\documentclass[prd,aps,groupedaddress,showpacs,amsmath]{revtex4}
\usepackage{epsf,epsfig,graphicx}

\newcounter{muni}

\begin{document}
\hbadness=10000 \pagenumbering{arabic}

\title{Gluonic contribution to $B\to \eta^{(\prime)}$ form factors}

\author{Yeo-Yie Charng$^{1}$}
\email{charng@phys.sinica.edu.tw}
\author{T. Kurimoto$^2$}
\email{krmt@sci.u-toyama.ac.jp}
\author{Hsiang-nan Li$^{1,3}$}
\email{hnli@phys.sinica.edu.tw}

\affiliation{$^{1}$Institute of Physics, Academia Sinica, Taipei,
Taiwan 115, Republic of China}
\affiliation{$^{2}$Department of
Physics, University of Toyama, Toyama 930-8555, Japan}
\affiliation{$^{3}$Department of Physics, National Cheng-Kung University,\\
Tainan, Taiwan 701, Republic of China}

\begin{abstract}

We calculate the flavor-singlet contribution to the
$B\to\eta^{(\prime)}$ transition form factors from the gluonic
content of the $\eta^{(\prime)}$ meson in the large-recoil region
using the perturbative QCD approach. The formulation for the
$\eta$-$\eta'$ mixing in the quark-flavor and singlet-octet
schemes is compared, and employed to determine the chiral
enhancement scales associated with the two-parton twist-3
$\eta^{(\prime)}$ meson distribution amplitudes. It is found that
the gluonic contribution is negligible in the $B\to\eta$ form
factors, and reaches few percents in the $B\to\eta'$ ones. Its
impact on the accommodation of the measured $B\to\eta^{(\prime)}
K$ branching ratios in the perturbative QCD and QCD-improved
factorization approaches is elaborated.

\end{abstract}


\maketitle

\section{INTRODUCTION}

It is still uncertain whether the flavor-singlet contributions to
$B$ meson decays into $\eta^{(\prime)}$ mesons play an essential
role. The flavor-singlet contributions to the $B\to\eta^{(\prime)}
K$ branching ratios, including those from the $b\to sgg$
transition \cite{EKP04}, from the spectator scattering
\cite{Du:1997hs,DYZ99}, and from the weak annihilation, have been
analyzed in the QCD-improved factorization (QCDF) approach
\cite{BN02}. However, at least the piece from the weak
annihilation can not be estimated unambiguously due to the
presence of the end-point singularities. The flavor-singlet
contribution to the $B\to\eta^{(\prime)}$ transition form factors
from the gluonic content of the $\eta^{(\prime)}$ meson is also
involved in the annihilation amplitudes. Though this contribution
seems to be most crucial among all the flavor-singlet pieces in
the $B\to\eta^{(\prime)} K$ decays, it has been parameterized and
varied arbitrarily \cite{BN02}. The form factors associated with
the decays $B\to\eta^{(\prime)}l^+l^-$ were handled in a similar
way recently \cite{CQ06}. A sizable gluonic content in the $\eta'$
meson has been indicated from a phenomenological analysis of the
relevant data \cite{K99}. All these previous studies motivate us
to make a more definite estimate of the gluonic contribution in
the $B\to\eta^{(\prime)}$ form factors.

In this paper we shall calculate the gluonic contribution to the
$B\to\eta^{(\prime)}$ form factors in the large-recoil region
using the perturbative QCD (PQCD) approach \cite{LY1,KLS,LUY}.
This approach is based on $k_T$ factorization theorem
\cite{LS,LL04}, so the end-point singularities do not exist. It
has been proposed to extract this gluonic contribution from the
measured $B\to\eta^{(\prime)}l\nu$ decay spectra \cite{KOY03},
which are, however, not available yet. To proceed with the
calculation, we need to specify a scheme for the $\eta$-$\eta'$
mixing. After comparing the quark-flavor basis \cite{FKS} and the
conventional singlet-octet basis, we adopt the former, in which
fewer two-parton twist-3 $\eta^{(\prime)}$ meson distribution
amplitudes are introduced. To reduce the theoretical uncertainty
from the distribution amplitudes, we employ the Gegenbauer
coefficients constrained by the data of both exclusive processes
\cite{FKS,KP,AS02}, such as the $\eta^{(\prime)}$ transition form
factors, and inclusive processes \cite{AP03}, such as
$\Upsilon(1S)\to\eta' X$. It will be shown that the gluonic
contribution is negligible in the $B\to\eta$ form factors, and
reaches few percents in the $B\to\eta'$ ones.

Whether the observed $B\to\eta K$ and $B\to\eta'K$ branching
ratios \cite{HFAG} can be accommodated simultaneously still
attracts a lot of attentions \cite{DKOZ2,HJL}. We shall elaborate
the impact of the gluonic contribution in the
$B\to\eta^{(\prime)}$ form factors on this issue in the QCDF and
PQCD frameworks. As noticed in \cite{BN02}, the gluonic
contribution increases the branching ratios $B(B\to\eta' K)$, but
decreases $B(B\to\eta K)$. Since the QCDF predictions for both
$B(B\to\eta' K)$ and $B(B\to\eta K)$ fall short compared with the
data \cite{BN02}, the gluonic contribution does not help. On the
contrary, there is much room for this contribution to play in
PQCD: the flavor-singlet amplitudes were not taken into account in
the earlier PQCD analysis of the $B\to\eta^{(\prime)} K$ decays
\cite{KS01}, whose predictions for $B(B\to\eta' K)$ [$B(B\to\eta
K)$] are lower (higher) than the measured values. If stretching
the gluonic contribution, it is likely to accommodate the data of
the $B\to\eta^{(\prime)} K$ branching ratios in PQCD.

In Sec.~II we compare the quark-flavor and singlet-octet schemes
for the $\eta$-$\eta'$ mixing, and obtain the chiral enhancement
scales associated with the two-parton twist-3 distribution
amplitudes in both cases. In Sec.~III we derive the factorization
formulas for the quark and gluonic contributions in the
$B\to\eta^{(\prime)}$ form factors, and perform the numerical
evaluation, together with a detailed uncertainty analysis. Our
results are then compared with those obtained in the literature.
The impact of the gluonic contribution on the accommodation of the
measured $B\to\eta^{(\prime)}K$ branching ratios is discussed.
Section IV is the conclusion.

\section{$\eta$-$\eta'$ MIXING and DISTRIBUTION AMPLITUDES}

For the $\eta$-$\eta'$ mixing, the conventional singlet-octet
basis and the quark-flavor basis \cite{FKS} have been proposed. In
the latter the $q\bar q\equiv (u\bar u+d\bar d)/\sqrt{2}$ and
$s\bar s$ flavor states, labelled by the $\eta_q$ and $\eta_s$
mesons, respectively, are defined. The physical states $\eta$ and
$\eta'$ are related to the flavor states through a single angle
$\phi$,
\begin{equation}\label{qs}
   \left( \begin{array}{c}
    |\eta\rangle \\ |\eta'\rangle
   \end{array} \right)
   = U(\phi)
   \left( \begin{array}{c}
    |\eta_q\rangle \\ |\eta_s\rangle
   \end{array} \right) \;,
\end{equation}
with the matrix,
\begin{equation}
U(\phi)=\left( \begin{array}{cc}
    \cos\phi & ~-\sin\phi \\
    \sin\phi & \phantom{~-}\cos\phi
   \end{array} \right)\;.
\end{equation}
It has been postulated \cite{FKS} that only two decay constants
$f_q$ and $f_s$ need to be introduced:
\begin{eqnarray}
   \langle 0|\bar q\gamma^\mu\gamma_5 q|\eta_q(P)\rangle
   &=& -\frac{i}{\sqrt2}\,f_q\,P^\mu \;,\nonumber \\
   \langle 0|\bar s\gamma^\mu\gamma_5 s|\eta_s(P)\rangle
   &=& -i f_s\,P^\mu \;,\label{deffq}
\end{eqnarray}
for the light quark $q=u$ or $d$. This postulation is based on the
assumption that the intrinsic $\bar qq$ ($\bar ss$) component is
absent in the $\eta_s$ ($\eta_q$) meson, ie., based on the OZI
suppression rule. The decay constants associated with the $\eta$
and $\eta'$ mesons:
\begin{eqnarray}\label{deffh}
   \langle 0|\bar q\gamma^\mu\gamma_5 q|\eta^{(\prime)}(P)\rangle
   &=& -\frac{i}{\sqrt2}\,f_{\eta^{(\prime)}}^q\,P^\mu \;,\nonumber \\
   \langle 0|\bar s\gamma^\mu\gamma_5 s|\eta^{(\prime)}(P)\rangle
   &=& -i f_{\eta^{(\prime)}}^s\,P^\mu \;,
\end{eqnarray}
are then related to $f_q$ and $f_s$ via the same mixing matrix,
\begin{equation}
  \left(\begin{array}{cc}
  f_\eta^q & f_\eta^s \\
   f_{\eta'}^q & f_{\eta'}^s \\
\end{array} \right)=U(\phi)
\left(\begin{array}{cc}
  f_q & 0 \\
  0 & f_s \\
\end{array} \right)\;.
\label{fpi}
\end{equation}

Employing the equation of motion,
\begin{equation}
   \partial_\mu(\bar q\gamma^\mu\gamma_5 q) = 2im_q\,\bar q\gamma_5 q
   +\frac{\alpha_s}{4\pi}\,G_{\mu\nu}\,\widetilde{G}^{\mu\nu}\;,
   \label{eom}
\end{equation}
and the one corresponding to the $s$ quark, where $G$ is the
field-strength tensor and $\widetilde{G}$ the dual field-strength
tensor,
one derives the relation \cite{FKS},
\begin{eqnarray}
M_{qs}^2=U^\dagger(\phi)M^2 U(\phi)\;.\label{matrix}
\end{eqnarray}
In the above expression the mass matrices are given by,
\begin{eqnarray}
M^2&=&\left(\begin{array}{cc}
  m_{\eta}^2 & 0 \\
  0 & m_{\eta'}^2 \\
\end{array} \right)\;,\nonumber\\
M_{qs}^2&=&\left(\begin{array}{cc} m_{qq}^2+(\sqrt{2}/f_q)\langle
0|\alpha_sG{\tilde G}/(4\pi)|\eta_q\rangle & (1/f_s)\langle
0|\alpha_sG{\tilde G}/(4\pi)|\eta_q\rangle \\
              (\sqrt{2}/f_q)\langle
0|\alpha_sG{\tilde G}/(4\pi)|\eta_s\rangle &
m_{ss}^2+(1/f_s)\langle 0|\alpha_sG{\tilde G}/(4\pi)|\eta_s\rangle
\\
\end{array}\right)\;,
\end{eqnarray}
with the abbreviations,
\begin{eqnarray}
m_{qq}^2&=&\frac{\sqrt{2}}{f_q}\langle 0|m_u\bar u i\gamma_5
u+m_d\bar d
i\gamma_5 d|\eta_q\rangle\;,\nonumber\\
m_{ss}^2&=&\frac{2}{f_s}\langle 0|m_s\bar s i\gamma_5
s|\eta_s\rangle\;.\label{mqq}
\end{eqnarray}
The above matrix elements define the chiral enhancement scales
associated with the two-parton twist-3 $\eta_q$ and $\eta_s$ meson
distribution amplitudes. Note that the axial U(1) anomaly
\cite{JB05} is the only source of the non-diagonal elements of
$M_{qs}$ in the quark-flavor basis, also a consequence of the
postulation that leads to Eq.~(\ref{deffq}).

We repeat the above formalism for the $\eta$-$\eta'$ mixing in the
singlet-octet basis, where the $(u\bar u+d\bar d+s\bar
s)/\sqrt{3}$ and $(u\bar u+d\bar d-2s\bar s)/\sqrt{6}$ states,
labelled by the $\eta_1$ and $\eta_8$ mesons, respectively, are
considered. From Eq.~(\ref{qs}) and the quark content of the
$\eta_1$ and $\eta_8$ mesons, we have the decompositions of the
$\eta$ and $\eta'$ meson states in the singlet-octet basis,
\begin{equation}
   \left( \begin{array}{c}
    |\eta\rangle \\ |\eta'\rangle
   \end{array} \right)
   = U(\theta)
   \left( \begin{array}{c}
    |\eta_8\rangle \\ |\eta_1\rangle
   \end{array} \right) \;,
\end{equation}
with the angle $\theta=\phi-\theta_i$, $\theta_i$ being the ideal
mixing angle associated with the matrix,
\begin{equation}
U(\theta_i)=\left(\begin{array}{cc}
  \frac{1}{\sqrt{3}} & -\frac{\sqrt{2}}{\sqrt{3}} \\
 \frac{\sqrt{2}}{\sqrt{3}} & \frac{1}{\sqrt{3}} \\
\end{array} \right)\;. \label{ideal}
\end{equation}
The decay constants defined in terms of the SU(3) flavor-singlet
and flavor-octet axial-vector currents,
\begin{equation}\label{axial}
\langle 0|J_{\mu 5}^i|\eta^{(\prime)}(P)\rangle =
-i\,f_{\eta^{(\prime)}}^i\,P^\mu \;,
\end{equation}
for $i=1$ or 8, are related to $f_q$ and $f_s$ through
\begin{equation}
  \left(\begin{array}{cc}
  f_\eta^8 & f_\eta^1 \\
   f_{\eta'}^8 & f_{\eta'}^1 \\
\end{array} \right)=U(\phi)
\left(\begin{array}{cc}
  f_q & 0 \\
  0 & f_s \\
\end{array} \right)U^\dagger(\theta_i)\;.
\label{f18}
\end{equation}
Compared to the conventional parametrization,
\begin{equation}
  \left(\begin{array}{cc}
  f_\eta^8 & f_\eta^1 \\
   f_{\eta'}^8 & f_{\eta'}^1 \\
\end{array} \right)=U_{81}
\left(\begin{array}{cc}
  f_8 & 0 \\
  0 & f_1 \\
\end{array} \right)\;,
\label{f81}
\end{equation}
with the matrix,
\begin{eqnarray}
U_{81}&=&\left(\begin{array}{cc}
  \cos\theta_8 & -\sin\theta_1 \\
  \sin\theta_8 & \cos\theta_1 \\
\end{array} \right)\;,
\end{eqnarray}
the decay constants of the $\eta_1$ and $\eta_8$ mesons, $f_1$ and
$f_8$, respectively, then connect to $f_q$ and $f_s$.

Following the similar procedure, we derive the version of
Eq.~(\ref{matrix}) in the singlet-octet basis,
\begin{eqnarray}
M_{81}^2=U^\dagger(\theta)M^2 U_{81}\;,\label{ma81}
\end{eqnarray}
with the mass matrix,
\begin{eqnarray}
M_{81}^2&=&\left(\begin{array}{cc} m_{88}^2 &
m_{18}^2+(\sqrt{3}/f_1)\langle
0|\alpha_sG{\tilde G}/(4\pi)|\eta_8\rangle \\
m_{81}^2 & m_{11}^2+(\sqrt{3}/f_1)\langle 0|\alpha_sG{\tilde
G}/(4\pi)|\eta_1\rangle
\\
\end{array}\right)\;,
\end{eqnarray}
and the abbreviations,
\begin{eqnarray}
m_{88}^2&=&\frac{2}{\sqrt{6}f_8}\langle 0|m_u\bar u i\gamma_5
u+m_d\bar d i\gamma_5 d-2m_s\bar s i\gamma_5
s|\eta_8\rangle\;,\nonumber\\
m_{18}^2&=&\frac{2}{\sqrt{3}f_1}\langle 0|m_u\bar u i\gamma_5
u+m_d\bar d i\gamma_5 d+m_s\bar s i\gamma_5 s|\eta_8\rangle\;,
\nonumber\\
m_{81}^2&=&\frac{2}{\sqrt{6}f_8}\langle 0|m_u\bar u i\gamma_5
u+m_d\bar d i\gamma_5 d-2m_s\bar s i\gamma_5
s|\eta_1\rangle\;,\nonumber\\
m_{11}^2&=&\frac{2}{\sqrt{3}f_1}\langle 0|m_u\bar u i\gamma_5
u+m_d\bar d i\gamma_5 d+m_s\bar s i\gamma_5 s|\eta_1\rangle\;.
\label{m281}
\end{eqnarray}
The above matrix elements define the chiral enhancement scales
associated with the two-parton twist-3 $\eta_1$ and $\eta_8$ meson
distribution amplitudes.

Note that the four hadronic matrix elements $m_{qq}^2$,
$m_{ss}^2$, $\langle 0|\alpha_s G{\tilde G}/(4\pi)|\eta_q\rangle$,
and $\langle 0|\alpha_s G{\tilde G}/(4\pi)|\eta_s\rangle$ are
fixed by Eq.~(\ref{matrix}) in the quark-flavor basis, given the
three inputs $f_q$, $f_s$, and $\phi$. However, $M_{81}^2$
contains six matrix elements, which can not be fixed completely by
Eq.~(\ref{ma81}). In fact, two OZI suppressed matrix elements
$\langle 0|m_u\bar u i\gamma_5 u+m_d\bar d i\gamma_5
d|\eta_s\rangle$ and $\langle 0|m_s\bar s i\gamma_5
s|\eta_q\rangle$ have been dropped in \cite{FKS}. In the
singlet-octet basis an approximation can be made to reduce the
number of matrix elements \cite{AF89,EF05}: the terms proportional
to the light quark masses $m_u$ and $m_d$ are negligible compared
with those proportional to $m_s$ in Eq.~(\ref{m281}), which then
becomes
\begin{eqnarray}
m_{88}^2&=&-\frac{4}{\sqrt{6}f_8}\langle 0|m_s\bar s i\gamma_5
s|\eta_8\rangle\;,\;\;\;\;
m_{18}^2=-\frac{f_8}{\sqrt{2}f_1}m_{88}^2\;,
\nonumber\\
m_{11}^2&=&\frac{2}{\sqrt{3}f_1}\langle 0|m_s\bar s i\gamma_5
s|\eta_1\rangle\;,\;\;\;\;
m_{81}^2=-\frac{\sqrt{2}f_1}{f_8}m_{11}^2\;. \label{m88s}
\end{eqnarray}

The three inputs $f_q$, $f_s$, and $\phi$ in Eq.~(\ref{matrix})
have been extracted from the data of the relevant exclusive
processes \cite{FKS},
\begin{eqnarray}
& &f_q=(1.07\pm 0.02)f_\pi\;,\;\;\;\;f_s=(1.34\pm
0.06)f_\pi\;,\;\;\; \phi=39.3^\circ\pm 1.0^\circ\;, \label{qsp}
\end{eqnarray}
which correspond, via Eqs.~(\ref{f18}) and (\ref{f81}), to
\cite{AS03}
\begin{eqnarray}
f_8\approx 1.26 f_\pi\;,\;\;\;\;f_1\approx 1.17
f_\pi\;,\;\;\;\;\theta_8\approx-21.2^o\;,\;\;\;\;\theta_1\approx-9.2^o\;,
\end{eqnarray}
in the singlet-octet basis. The above values are well consistent
with those in \cite{EF05}, which were also extracted from the
relevant data but based on the approximation in Eq.~(\ref{m88s}).
We stress that the approximations in the quark-flavor basis
\cite{FKS} and in the singlet-octet basis \cite{EF05} for
decreasing the number of hadronic matrix elements are very
different. Therefore, the above agreement implies that these
approximations make sense, and that the two bases will be
equivalent to each other, if one chooses the parameters obeying
the constraints in Eqs.~(\ref{matrix}), (\ref{f18}), (\ref{f81}),
(\ref{ma81}), and (\ref{m88s}) for a calculation.

Viewing Eqs.~(\ref{mqq}) and (\ref{m281}), it is obvious that
fewer two-parton twist-3 distribution amplitudes are introduced in
the quark-flavor scheme \footnote{The simplicity of the
quark-flavor scheme is lost at higher orders, since the evolution
effect will cause the mixing between the $\eta_q$ and $\eta_s$
meson distribution amplitudes.}. Hence, we adopt this scheme for
the $\eta$-$\eta'$ mixing, and specify the $\eta_{q}$ and
$\eta_{s}$ meson distribution amplitudes. Their two-parton quark
components are defined via the nonlocal matrix elements,
\begin{eqnarray}
\langle\eta_q(P)|{\bar q}_{\gamma}^a(z)q_{\beta}^b(0)|0\rangle
&=&-\frac{i}{2\sqrt{N_c}} \delta^{ab}\int_0^1dx e^{ixP\cdot
z}\left\{[\gamma_5\not P]_{\beta\gamma}\phi^A_q(x) +
[\gamma_5]_{\beta\gamma}m_0^q\phi^P_q(x)\right.
\nonumber\\
& &\left.+m_0^q[\gamma_5(\not n_+\not n_--1)]_{\beta\gamma}
\phi^T_q(x)\right\}\;,\nonumber\\
\langle\eta_s(P)|{\bar s}_{\gamma}^a(z)s_{\beta}^b(0)|0\rangle
&=&-\frac{i}{\sqrt{2N_c}}\delta^{ab} \int_0^1dx e^{ixP\cdot
z}\left\{[\gamma_5\not P]_{\beta\gamma}\phi^A_s(x) +
[\gamma_5]_{\beta\gamma}m_0^s\phi^P_s(x)\right.
\nonumber\\
& &\left.+m_0^s[\gamma_5(\not n_+\not n_--1)]_{\beta\gamma}
\phi^T_s(x)\right\}\;, \label{fs}
\end{eqnarray}
where $P=(P^+,0,{\bf 0}_T)$ is the $\eta_{q,s}$ meson momentum,
the light-like vector $z=(0,z^-,{\bf 0}_T)$ the coordinate of the
$q$ and $s$ quarks, the dimensionless vector $n_+=(1,0,{\bf 0}_T)$
parallel to $P$, $n_-=(0,1,{\bf 0}_T)$ parallel to $z$, the
superscripts $a$ and $b$ the color indices, the subscripts
$\gamma$ and $\beta$ the Dirac indices, $N_c=3$ the number of
colors, and $x$ the momentum fraction carried by the $q$ and $s$
quarks. The chiral enhancement scales $m_0^{q}$ and $m_0^{s}$ have
been fixed by Eq.~(\ref{matrix}), whose explicit expressions are
\begin{eqnarray}
m_0^q&\equiv& \frac{m_{qq}^2}{2m_q}=\frac{1}{2m_q}
\left[m_{\eta}^2\cos^2\phi+m_{\eta'}^2\sin^2\phi-
\frac{\sqrt{2}f_s}{f_q}(m_{\eta'}^2-m_\eta^2)\cos\phi\sin\phi\right]\;,\nonumber\\
m_0^s&\equiv& \frac{m_{ss}^2}{2m_s}=\frac{1}{2m_s}
\left[m_{\eta'}^2\cos^2\phi+m_{\eta}^2\sin^2\phi-
\frac{f_q}{\sqrt{2}f_s}(m_{\eta'}^2-m_\eta^2)\cos\phi\sin\phi\right]\;,
\end{eqnarray}
respectively, assuming the exact isospin symmetry $m_q\equiv
m_u=m_d$ (For the inclusion of the isospin symmetry breaking
effect, refer to \cite{K05}).

The distribution amplitudes $\phi_{q,s}^A$ are twist-2, and
$\phi_{q,s}^P$ and $\phi_{q,s}^T$ twist-3. As explained in
\cite{TLS}, both the twist-2 and twist-3 two-parton distribution
amplitudes contribute at leading power in the analysis of
exclusive $B$ meson decays. We follow the parametrization for the
pion distribution amplitudes proposed in \cite{Ball:2004ye},
\begin{eqnarray}
\phi^A_{q(s)}(x) &=& \, \frac{f_{q(s)}}{2\sqrt{2N_c}} 6x(1-x) \left[1 +
a_2^{q(s)}C_2^{3/2}(2x-1)\right] \;,
\nonumber\\
\phi^P_{q(s)}(x) &=&  \, \frac{f_{q(s)}}{2\sqrt{2N_c}} \bigg[ 1 +\left(30\eta_3
-\frac{5}{2}\rho_{q(s)}^2\right) C_2^{1/2}(2x-1)
\nonumber\\
& & \hspace{15mm} -\, 3\left\{ \eta_3\omega_3 +
\frac{9}{20}\rho_{q(s)}^2(1+6a_2^{q(s)}) \right\}
C_4^{1/2}(2x-1) \bigg]\;,\nonumber\\
\phi^T_{q(s)}(x) &=& \, \frac{f_{q(s)}}{2\sqrt{2N_c}} (1-2x)\bigg[ 1 + 6\left(5\eta_3
-\frac{1}{2}\eta_3\omega_3 - \frac{7}{20}
      \rho_{q(s)}^2 - \frac{3}{5}\rho_{q(s)}^2 a_2^{q(s)} \right)
(1-10x+10x^2) \bigg]\;,\label{qda}
\end{eqnarray}
with the mass ratios $\rho_{q}=2m_q/m_{qq}$ and
$\rho_{s}=2m_{s}/m_{ss}$, and the Gegenbauer polynomials,
\begin{eqnarray}
C_2^{1/2}(t)\, =\, \frac{1}{2} \left(3\, t^2-1\right)
\;,\;\;\;\;C_2^{3/2}(t)\, =\, \frac{3}{2} \left(5\, t^2-1\right)
\;,\;\;\;\; C_4^{1/2}(t)\, =\, \frac{1}{8} \left(3-30\, t^2+35\,
t^4\right) \;.
\end{eqnarray}
The values of the constant parameters $a_2^q$, $\eta_3$ and
$\omega_3$ in Eq.~(\ref{qda}) will be given in the next section.

The leading-twist gluonic distribution amplitudes of the $\eta_q$
and $\eta_s$ mesons are defined by \cite{AP03}
\begin{eqnarray}
\langle\eta_q(P)|A_{[\mu}^a(z)A_{\nu]}^b(0)|0\rangle
&=&\frac{\sqrt{2}f_q}{\sqrt{3}}\frac{C_F}{4\sqrt{3}}
\frac{\delta^{ab}}{N_c^2-1}\epsilon_{\mu\nu\rho\sigma}\,
\frac{n_-^\rho P^\sigma\,}{n_-\cdot P} \int_0^1dx
e^{ixP\cdot z}\frac{\phi_{q}^{G}(x)}{x(1-x)} \;,\nonumber\\
\langle\eta_s(P)|A_{[\mu}^a(z)A_{\nu]}^b(0)|0\rangle
&=&\frac{f_s}{\sqrt{3}}\frac{C_F}{4\sqrt{3}}
\frac{\delta^{ab}}{N_c^2-1}\epsilon_{\mu\nu\rho\sigma}\,
\frac{n^\rho P^\sigma\,}{n\cdot P} \int_0^1dx e^{ixP\cdot
z}\frac{\phi_{s}^{G}(x)}{x(1-x)} \;, \label{qgn}
\end{eqnarray}
with the notation $A_{[\mu}^a(z)A_{\nu]}^b(w)\equiv
[A_{\mu}^a(z)A_{\nu}^b(w)-A_{\nu}^a(z)A_{\mu}^b(w)]/2$ and the
function \cite{KP,Terentev:qu,Ohrndorf:uz,Shifman:dk,Baier:pm},
\begin{eqnarray}
\phi_{q(s)}^{G}(x)=x^2(1-x)^2
B_2^{q(s)}C_1^{5/2}(2x-1)\;,\;\;\;\;C_1^{5/2}(t)=5t\;.
\label{phiG}
\end{eqnarray}
According to the above definition, the gluon labelled by the
subscript $\mu$ carries the fractional momentum $xP$. The two
Gegenbauer coefficients $B_2^q$ and $B_2^s$ could be different in
principle. However, due to the large uncertainty in their values,
it is acceptable to assume $B_2^q=B_2^s\equiv B_2$. As shown
later, the contribution from the above gluonic distribution
amplitudes is smaller than that from the quark distribution
amplitudes. Therefore, the subleading-twist gluonic distribution
amplitudes of the $\eta_{q,s}$ meson will be dropped below.

The gluonic distribution amplitude of the $\eta'$ meson defined in
\cite{Ali03} is related to those in Eq~(\ref{qgn}) through
Eq.~(\ref{qs}),
\begin{equation}
   \left( \begin{array}{c}
    \langle\eta|A_{[\mu}^a(z)A_{\nu]}^b(0)|0\rangle \\
    \langle\eta'|A_{[\mu}^a(z)A_{\nu]}^b(0)|0\rangle
   \end{array} \right)
   = U(\phi)
   \left( \begin{array}{c}
    \langle\eta_q|A_{[\mu}^a(z)A_{\nu]}^b(0)|0\rangle \\
    \langle\eta_s|A_{[\mu}^a(z)A_{\nu]}^b(0)|0\rangle
   \end{array} \right) \;,
\label{geta}
\end{equation}
which also defines the gluonic distribution amplitude of the
$\eta$ meson. Besides, our parametrization of the gluonic
distribution amplitudes are identical to those in \cite{BN02}
except a different definition of $B_2$: their Gegenbauer
coefficient is 2/9 times of ours.

\section{$B\to\eta^{(\prime)}$ FORM FACTORS}

After defining the quark and gluonic distribution amplitudes of
the $\eta_q$ and $\eta_s$ mesons, we are ready to calculate the
$B\to\eta^{(\prime)}$ transition form factors at leading order of
the strong coupling constant $\alpha_s$. In the $B$ meson rest
frame, we choose the $B$ meson momentum $P_1$ and the
$\eta^{(\prime)}$ meson momentum $P_2$ in the light-cone
coordinates:
\begin{eqnarray}
P_1=\frac{m_B}{\sqrt{2}}(1,1,{\bf 0}_T)\;,\;\;\;
P_2=\frac{m_B}{\sqrt{2}}(\rho,0,{\bf 0}_T)\;, \label{pa}
\end{eqnarray}
where the energy fraction $\rho$ carried by the $\eta^{(\prime)}$
meson is related to the lepton-pair momentum $q=P_1-P_2$ via
$q^2=(1-\rho)m_B^2$, $m_B$ being the $B$ meson mass. The
$\eta^{(\prime)}$ meson mass, appearing only in power-suppressed
terms, has been neglected. The spectator momenta $k_1$ on the $B$
meson side and $k_2$ on the $\eta^{(\prime)}$ meson side are
parameterized as
\begin{eqnarray}
k_1=\left(0,x_1\frac{m_B}{\sqrt{2}},{\bf k}_{1T}\right)\;,\;\;\;
k_2=\left(x_2\rho\frac{m_B}{\sqrt{2}},0,{\bf k}_{2T}\right)\;,
\label{k12}
\end{eqnarray}
$x_1$ and $x_2$ being the parton momentum fractions, and ${\bf
k}_{1T}$ and ${\bf k}_{2T}$ the parton transverse momenta.

We first compute the $B\to\eta_{q(s)}$ form factors defined by the
local matrix elements,
\begin{eqnarray}
\langle \eta_{q(s)}(P_2)|{\bar b}(0)\gamma_\mu u(0)|B(P_1)\rangle
&=&F_{+}^{B\eta_{q(s)}}(q^2)\left[(P_1+P_2)_\mu
-\frac{m_B^2-m_{\eta_{q(s)}}^2}{q^2}q_\mu\right]\nonumber\\
& &+F_{0}^{B\eta_{q(s)}}(q^2)
\frac{m_B^2-m_{\eta_{q(s)}}^2}{q^2}q_\mu\;,\nonumber\\
\langle \eta_{q(s)}(P_2)|{\bar b}(0) i\sigma^{\mu\nu} q_\nu
d(0)|B(P_1) \rangle &=& \frac{
F_T^{B\eta_{q(s)}}(q^2)}{m_B+m_{\eta_{q(s)}}}\left[
(m_B^2-m_{\eta_{q(s)}}^2)\,q^\mu-q^2(P_1^\mu+P_2^{\mu})\right]\;,
\label{ftensor}
\end{eqnarray}
where the $\eta_{q(s)}$ meson mass $m_{\eta_{q(s)}}$ will be set
to zero eventually. The form factors $F_{+,0}$ are associated with
the semileptonic decay $B\to\eta^{(\prime)}l\nu$, and $F_T$ with
$B\to\eta^{(\prime)}l^+l^-$. For the involved $\bar b\to \bar
u,\bar d$ transitions, the above form factors are decomposed into
\begin{eqnarray}
& &F_{+(0,T)}^{B\eta_{q}}=F_{q+(0,T)}^{B\eta_{q}}
+F_{g+(0,T)}^{B\eta_{q}}\;,\;\;\;\;
F_{+(0,T)}^{B\eta_{s}}=F_{g+(0,T)}^{B\eta_{s}}\;.
\end{eqnarray}
That is, the $\eta_s$ meson state contributes only through the
flavor-singlet pieces $F_{g+,g0,gT}^{B\eta_{s}}$. The
$B\to\eta^{(\prime)}$ form factors are then obtained from the
mixing,
\begin{eqnarray}
   \left( \begin{array}{c}
    F^{B\eta}_{+(0,T)} \\
    F^{B\eta'}_{+(0,T)}
   \end{array} \right)
   = U(\phi)
   \left( \begin{array}{c}
    F^{B\eta_q}_{+(0,T)} \\
    F^{B\eta_s}_{+(0,T)}
   \end{array} \right) \;.\label{betap}
\end{eqnarray}
It is then expected that the gluonic contribution is more
significant in the $B\to\eta'$ form factors than in the $B\to\eta$
ones, since those from the $\eta_q$ and $\eta_s$ mesons add up in
the former, but partially cancel in the latter.

The factorization formulas for $F_{q+,q0,qT}^{B\eta_{q}}$ are
similar to those for the $B\to\pi$ form factors \cite{TLS,WY}:
\begin{eqnarray}
F_{q+}^{B\eta_{q}}(q^2)&=&\frac{8}{\sqrt{2}}\pi
m_B^2 C_F\int dx_1dx_2\int b_1db_1
b_2db_2\phi_B(x_1,b_1)
\nonumber\\
&\times& \Biggl\{\left[(1+x_2\rho)\phi_q^A(x_2) +r_q
\left(\frac{2}{\rho}-1-2x_2\right)\phi_q^T(x_2)+r_q(1-2x_2)\phi_q^P(x_2)
\right]E(t^{(1)})h(x_1,x_2,b_1,b_2)
\nonumber\\
& &+ 2r_q\phi_q^P E(t^{(2})h(x_2,x_1,b_2,b_1)\Biggr\}\;,\nonumber\\
F_{q0}^{B\eta_{q}}(q^2)&=&\frac{8}{\sqrt{2}}\pi
m_B^2 C_F\rho\int dx_1dx_2\int b_1db_1
b_2db_2\phi_B(x_1,b_1)
\nonumber\\
&\times& \Biggl\{\left[(1+x_2\rho)\phi_q^A(x_2) +r_q
\left(1-2x_2\right)\phi_q^T(x_2)+r_q\left(\frac{2}{\rho}-1-2x_2\right)\phi_q^P(x_2)
\right]E(t^{(1)})h(x_1,x_2,b_1,b_2)
\nonumber\\
& &+ 2r_q\phi_q^P E(t^{(2})h(x_2,x_1,b_2,b_1)\Biggr\}\;,
\nonumber\\
F_{qT}^{B\eta_{q}}(q^2)&=&\frac{8}{\sqrt{2}}\pi
m_B^2 C_F\int dx_1dx_2\int b_1db_1
b_2db_2\phi_B(x_1,b_1)
\nonumber\\
&\times& \Biggl\{\left[\phi_q^A(x_2) +r_q
\left(\frac{2}{\rho}+x_2\right)\phi_q^T(x_2)- r_qx_2\phi_q^P(x_2)
\right]E(t^{(1)})h(x_1,x_2,b_1,b_2)
\nonumber\\
& &+ 2r_q\phi_q^P E(t^{(2})h(x_2,x_1,b_2,b_1)\Biggr\}\;, \label{qfpic}
\end{eqnarray}
with the color factor $C_F=4/3$, the $B$ meson wave function
$\phi_B$, the impact parameter $b_1$ ($b_2$) conjugate to the
parton transverse momentum $k_{1T}$ ($k_{2T}$), the mass ratio
$r_q=m_0^q/m_B$, the hard function $h$, and the evolution factor,
\begin{eqnarray}
E(t)=\alpha_s(t)e^{-S_B(t)-S_{\eta_q}(t)}\;. \label{evol}
\end{eqnarray}
The choice of the hard scale $t$, and the explicit expressions for
$h$, for the Sudakov exponent $S_B$ associated with the $B$ meson,
and for the Sudakov exponent $S_{\eta_q}$ associated with a light
meson are referred to \cite{TLS}. The threshold resummation factor
associated with the hard function is the same as in the $B\to\pi$
form factors \cite{UL}. The coefficient $1/\sqrt{2}$ appears,
because only the $u$ or $d$ quark component of the $\eta_q$ meson
is involved. We point out that the term proportional to $2/\rho$
in $F_{qT}^{B\eta_q}$ is missed in \cite{CG02}.

\begin{figure}[t]
\begin{center}
\includegraphics[scale=0.6]{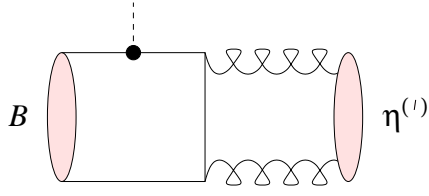}
\caption{Gluonic contribution to the $B\to\eta^{(\prime)}$ form
factors. Another diagram with the two gluons crossed is
suppressed.}\label{fig:ff}
\end{center}
\end{figure}

For the gluonic contribution, it has been argued \cite{BN02} that
the diagram in Fig.~\ref{fig:ff} with the two gluon emitted from
the light spectator quark of the $B$ meson is leading. Another
diagram with the two gluons crossed, giving the identical
contribution, is not displayed. The third diagram vanishes, in
which the virtual gluon from the $u$ and $\bar u$ quark
annihilation couples the two valence gluons in the
$\eta^{(\prime)}$ meson. The flavor-singlet pieces in the
$B\to\eta_{q}$ form factors are written as
\begin{eqnarray}
F_{g+}^{B\eta_{q}}(q^2)&=&-\frac{8}{3}\pi
m_B^2 f_q\frac{C_F^2\sqrt{N_c}}{N_c^2-1}\int dx_1dx_2\int b_1db_1
b_2db_2\phi_B(x_1,b_1)\frac{\phi^G(x_2)}{x_2(1-x_2)}
\nonumber\\
& &\times x_1[1+(1-\rho)x_2)]E(t^{(2})h(x_2,x_1,b_2,b_1)\;,\nonumber\\
F_{g0}^{B\eta_{q}}(q^2)&=&-\frac{8}{3}\pi
m_B^2 f_q\frac{C_F^2\sqrt{N_c}}{N_c^2-1}\rho\int dx_1dx_2\int b_1db_1
b_2db_2\phi_B(x_1,b_1)\frac{\phi^G(x_2)}{x_2(1-x_2)}
\nonumber \\
& &\times x_1[1-(1-\rho)x_2)]E(t^{(2})h(x_2,x_1,b_2,b_1)\;,\nonumber\\
F_{gT}^{B\eta_{q}}(q^2)&=&-\frac{8}{3}\pi
m_B^2 f_q\frac{C_F^2\sqrt{N_c}}{N_c^2-1}\int dx_1dx_2\int b_1db_1
b_2db_2\phi_B(x_1,b_1)\frac{\phi^G(x_2)}{x_2(1-x_2)}
\nonumber\\
& &\times x_1(1+x_2)E(t^{(2})h(x_2,x_1,b_2,b_1)\;. \label{gfpi}
\end{eqnarray}
A factor 2 has been included, which is attributed to the identical
contribution from the second diagram mentioned above. The
calculation is similar to that of the $\eta' g^* g^{(*)}$ vertex
in \cite{Muta:1999tc,Ali:2000ci}. The expressions for
$F_{g+,g0,gT}^{B\eta_{s}}$ are the same as in Eq.~(\ref{gfpi}),
but with $f_q$ being replaced by $f_s/\sqrt{2}$. It is observed
that Eq.~(\ref{gfpi}) is proportional to the small momentum
fraction $x_1\sim \Lambda/m_B$ \cite{CL0505}, $\Lambda$ being a
hadronic scale, compared to the quark contribution.
Because the above factorization formulas do not develop the
end-point singularities, if removing $k_T$, the gluonic
contribution can in fact be computed in collinear factorization
theorem.

The evolution factor in Eq.~(\ref{gfpi}) is given by
\begin{eqnarray}
E(t)=\alpha_s(t)e^{-S_B(t)-S_G(t)}\;, \label{gvol}
\end{eqnarray}
where the Sudakov exponent $S_G$ is associated with the gluonic
distribution amplitudes of the $\eta_{q,s}$ mesons. Following the
studies in \cite{LL99}, its expression is, up to the
leading-logarithm accuracy, similar to $S_{\eta_q}$, but with the
anomalous dimension $\alpha_sC_F/\pi$ being replace by
$\alpha_sC_A/\pi$, $C_A=3$ being a color factor:
\begin{eqnarray}
S_G(t)&=&s_G(x_2P_2^+,b_2)+s_G((1-x_2)P_2^+,b_2)\;,\nonumber\\
s_G(Q,b)&=&\int_{1/b}^{Q}\frac{d
\mu}{\mu}\left[\ln\left(\frac{Q}{\mu} \right)A(\alpha_s(\mu))
\right]\;,\;\;\;\; A=\frac{\alpha_s}{\pi}C_A\;.
\end{eqnarray}
That is, the Sudakov suppression is stronger in the gluonic
distribution amplitudes than in the quark ones. There is no point
to include the next-to-leading-logarithm resummation, since the
gluonic distribution amplitudes are not very certain yet. For
consistency, we neglect the single-logarithm renormalization-group
summation governed by the anomalous dimension of the gluon wave
function. We adopt the one-loop expression of the running coupling
constant $\alpha_s$, when evaluating the above Sudakov factors.

We then perform a detailed numerical analysis, including that of
theoretical uncertainties \cite{Kuri06}. The $B$ meson wave
function is the same as in \cite{LMS05} with the shape parameter
varied between $\omega_B=(0.40\pm 0.04)$ GeV. There is another $B$
meson wave function in the heavy-quark limit \cite{GN}, whose
contribution to transition form factors may be finite. However, it
has been shown that its effect can be well mimicked by a single
$B$ meson wave function, if a suitable $\omega_B$ is chosen
\cite{Kuri06}. Therefore, we adopt this approximation, and vary
$\omega_B$ in the above range. To obtain the chiral enhancement
scale $m_0^q$, we need the masses $m_\eta=0.548$ GeV and
$m_{\eta'}=0.958$ GeV, and the inputs in Eq.~(\ref{qsp}). Because
meson distribution amplitudes are defined at 1 GeV, we take the
light quark mass $m_q(1\;{\rm GeV})=(5.6\pm 1.6)$ MeV
\cite{BBL0603}. We have confirmed that the twist-2 and twist-3
quark distribution amplitudes of the $\eta'$ meson in \cite{Ali03}
are consistent with Eqs.~(\ref{ma81}) and (\ref{fs}), if their
$\eta'$ meson decay constant $f_{\eta'}$ is regarded as
$f_{\eta'}^1$ in the singlet-octet basis. Hence, it is legitimate
to adopt $a_2^{q}=-0.008\pm 0.054$ extracted from the relevant
data \cite{AP03} for the twist-2 distribution amplitude. The
twist-3 distribution amplitudes have not yet been constrained
experimentally, so we choose $\eta_3=0.015$ and $\omega_3=-3$ the
same as for the pion distribution amplitudes \cite{Ball:2004ye}.
It is not necessary to specify the parameters involved in the
quark distribution amplitudes of the $\eta_s$ meson here. The
overall coefficients in Eq.~(\ref{qgn}) have been arranged in the
way that the decay constants in \cite{Ali03} and in
Eq.~(\ref{qgn}) satisfy Eq.~(\ref{f18}). Following
Eq.~(\ref{geta}), the range of $B_2=4.6\pm 2.5$ extracted in
\cite{AP03} can also be adopted directly. The theoretical
uncertainty arising from the variation of the above parameters
will be investigated.

\begin{table}[ht]
\begin{center}
\begin{tabular}{c|cccc}\hline
\rule{0mm}{4mm} & $F^{B\eta}_{+,0}$ & $F^{B\eta}_T$
 & $F^{B\eta'}_{+,0}$ & $F^{B\eta'}_T$\\\hline
$F(0)$ & 0.147 & 0.139 & 0.121 & 0.114\\
ratio & 0.0031& 0.0028 & 0.023  & 0.021\\
$F_1$ & 0.252 & 0.237 & 0.252 & 0.237\\
$F_2$ & 0.0034 & 0.0029 & 0.0034 & 0.0029\\\hline
\end{tabular}
\caption{%
Form factor values at maximal recoil, the ratios of their gluonic
contributions, and values of $F_{1,2}$ defined in Eq~(\ref{40}).
}\label{tbl1}
\end{center}
\end{table}

The following parametrization for the $B\to\eta^{(\prime)}$ form
factors was proposed in \cite{BN02},
\begin{equation}
F_0^{B\eta^{(\prime)}}(0)=
F_1\,\frac{f_{\eta^{(\prime)}}^q}{\sqrt{2}f_\pi} +
F_2\,\frac{\sqrt2
f_{\eta^{(\prime)}}^q+f_{\eta^{(\prime)}}^s}{\sqrt6
f_\pi}\;,\label{40}
\end{equation}
where $F_1$ ($F_2$) corresponds to the quark (gluonic)
contribution, and a factor $1/\sqrt{2}$ is included to match our
convention. $F_1$ has been set to the $B\to\pi$ form factor, and
the unknown $F_2$ varied arbitrarily between $[0,0.1]$
\cite{BN02}. The above parametrization also applies to the other
form factors $F_+$ and $F_T$. It is easy to identify, from
Eq.~(\ref{betap}), the relations of $F_{1,2}$ to our form factors,
\begin{equation}
F_1 = \frac{\sqrt{2}f_\pi}{f_q} F_q^{B\eta_q}(0)\;, \ \ F_2 =
\frac{\sqrt{3}f_\pi}{f_q} F_g^{B\eta_q}(0)\;. \label{f12par}
\end{equation}
To have a picture of the magnitude of the gluonic contribution, we
present in Table~\ref{tbl1} the central values of the form factors
at maximal recoil, the ratios of their gluonic contributions, and
the values of $F_{1,2}$. It is indicated that the gluonic
contribution is about 0.3\% in the $B\to\eta$ form factors, and
about 2\% in the $B\to\eta'$ ones. The central value of $F_1$ for
the form factors $F_{+,0}^{B\eta^{(\prime)}}$ is indeed close to
the $B\to\pi$ form factors $F^{B\pi}_{+,0}(0)$ \cite{TLS}. The
central value of $F_2=0.0034$ for $F_{+,0}^{B\eta^{(\prime)}}$ is
located within $[0,0.1]$ \cite{BN02}, but near the lower end of
the interval (to be precise, $F_2$ runs between $[0.0016,0.0052]$,
considering the range of the parameter $B_2$).
We notice the difference between $F_2$ for
$F_{+,0}^{B\eta^{(\prime)}}(0)$ and $F_2$ for
$F_{T}^{B\eta^{(\prime)}}(0)$, which is obvious from the
factorization formulas in Eq.~(\ref{gfpi}): $F_{gT}^{B\eta_q}(0)$
contains an additional term proportional to $x_2$ compared with
$F_{g+,g0}^{B\eta_q}(0)$. Our observation is thus contrary to the
assumption $F_{g+,g0}^{B\eta_q}(0)=F_{gT}^{B\eta_q}(0)$ postulated
in the analysis of the $B\to\eta^{(\prime)}l^+l^-$ decays
\cite{CQ06}. Nevertheless, this difference is not crucial for the
estimate in \cite{CQ06}, because the gluonic contribution is not
dominant.

\begin{figure}[ht]
\begin{center}
\resizebox{6cm}{!}{\includegraphics{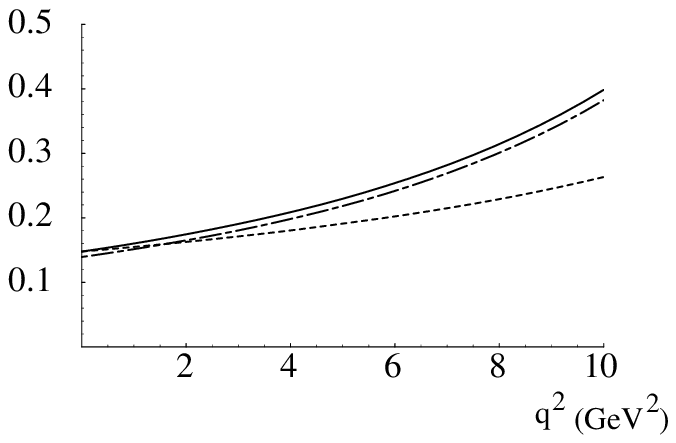}} \ \
\resizebox{6cm}{!}{\includegraphics{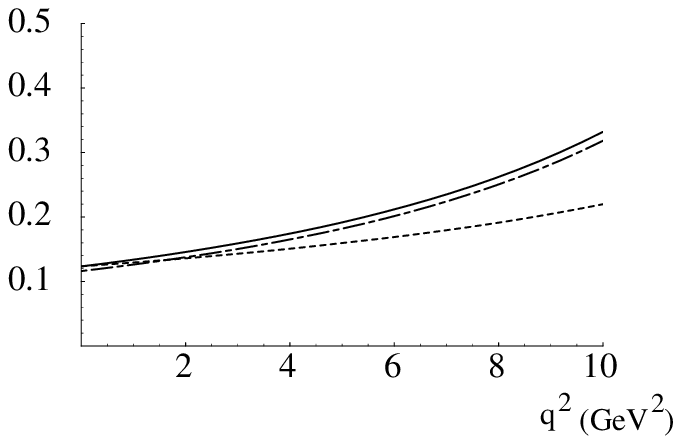}} \ \

(a)\hskip6cm (b)

\caption{$q^2$ dependence of (a) $F^{B\eta}_{+,0,T}$ and (b)
$F^{B\eta'}_{+,0,T}$, corresponding to the central values of the
inputs with the solid, dotted and dash-dotted lines representing
the $F_+$, $F_0$, and $F_T$, respectively.}\label{feta}
\end{center}
\end{figure}
\begin{figure}[ht]
\begin{center}
\resizebox{6cm}{!}{\includegraphics{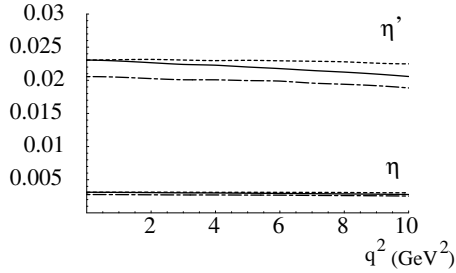}} \ \

\caption{$q^2$ dependence of the ratio of the gluonic contribution
to the total values of $F^{B\eta(\eta')}_+$ (solid lines),
$F^{B\eta(\eta')}_0$ (dotted lines), and $F^{B\eta(\eta')}_T$
(dash-dotted lines) corresponding to the central values of the
inputs. }\label{fratio}
\end{center}
\end{figure}

The $q^2$ dependence of the $B\to\eta$ and $B\to\eta'$ form
factors corresponding to the central values of the inputs are
displayed in Figs.~\ref{feta} (a) and (b). The ratios of the
gluonic contributions to the total values of the form factors are
shown in Fig.~\ref{fratio}. It is found that the gluonic
contribution remains negligible in the $B\to\eta$ form factors,
and about 2\% in the $B\to\eta'$ ones in the whole large-recoil
region. The form factors $F^{B\eta^{(\prime)}}_0(q^2)$ exhibit a
smaller slope with $q^2$, because the overall coefficients of
$F^{B\eta_q}_{q0,g0}$ contain the energy fraction $\rho$ as shown
in Eqs.~(\ref{qfpic}) and (\ref{gfpi}), consistent with the
large-energy form-factor relation in \cite{BF}. The $B\to\eta$
form factors are proportional to $\cos\phi F^{B\eta_q}_q$, and the
$B\to\eta'$ form factors to $\sin\phi F^{B\eta_q}_q$ plus some
amount of gluonic contributions. Because of the small gluonic
contribution indicated in Fig.~\ref{fratio}, we have
$F^{B\eta}(q^2)>F^{B\eta'}(q^2)$ shown in Fig.~\ref{feta} simply
due to $\cos\phi >\sin\phi$ for $\phi\approx 39.3^o$. This
observation is in agreement with the tendency exhibited in the
measurement of the semileptonic branching ratios \cite{BaBar},
\begin{eqnarray}
B(B^+\to\eta l^+\nu_l)&=&(0.84\pm 0.27\pm 0.21)\times
10^{-4}<1.4\times 10^{-4}\; (90\%\; {\rm C.L.})\;,\nonumber\\
B(B^+\to\eta' l^+\nu_l)&=&(0.33\pm 0.60\pm 0.30)\times
10^{-4}<1.3\times 10^{-4}\; (90\%\; {\rm C.L.})\;.
\end{eqnarray}
More precise data will provide the information on the importance
of the gluonic contribution. Our analysis implies that the $q^2$
dependence of the gluonic contribution is weaker than that of the
quark contribution. Hence, the assumption \cite{CQ06,KOY03} that
both pieces show the same $q^2$ dependence is not appropriate.

\begin{figure}[ht]
\begin{center}
\resizebox{5cm}{!}{\includegraphics{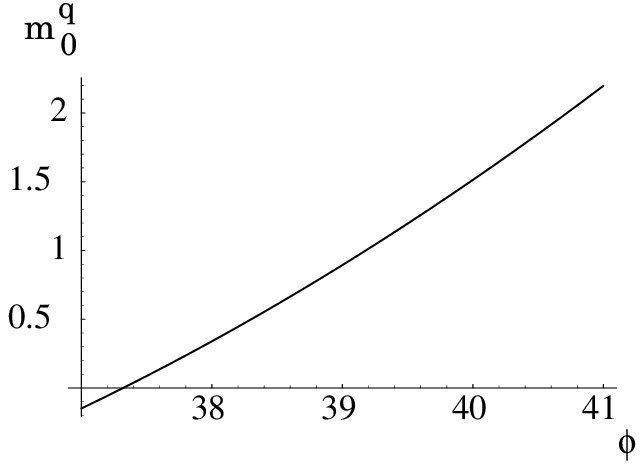}} \ \hskip3cm
\resizebox{5cm}{!}{\includegraphics{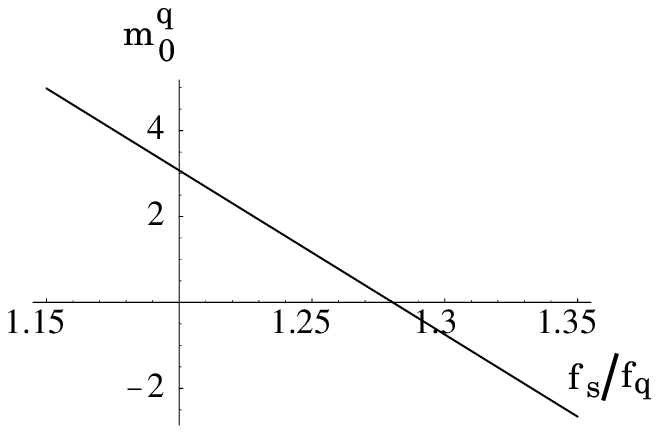}} \ \

(a)\hskip8cm (b)

\caption{Dependence of $m_0^q$ (a) on $\phi$ in units of degrees,
and (b) on $f_s/f_q$.}\label{figm0}
\end{center}
\end{figure}

\begin{table}[ht]
\begin{center}
\begin{tabular}{l|cccc}\hline
\rule{0mm}{4mm} & $F^{B\eta}_{+,0}(0)$ & $F^{B\eta}_T(0)$
 & $F^{B\eta'}_{+,0}(0)$ & $F^{B\eta'}_T(0)$\\\hline
$\omega_B=$ (0.36 - 0.44)&
0.174 - 0.127 &0.164 - 0.119 &0.127 - 0.106 &0.136 - 0.099 \\
$f_q=$ (1.05 - 1.09)$f_\pi$&
0.057 - 0.239  &0.049 - 0.230  &0.049 - 0.198  & 0.042 - 0.190 \\
$f_s=$ (1.28 - 1.40)$f_\pi$& 0.359 - (--0.063)  & 0.348 -
(--0.070)  & 0.296 - (--0.049)  & 0.287 - (--0.055)   \\
$\phi=$ (38.3 - 40.3)&
0.093 - 0.208  & 0.084 - 0.199  & 0.076 - 0.179  & 0.069 - 0.171  \\
$m_q=$ (4.0 - 7.2)&
0.191 - 0.124  & 0.181 - 0.116  & 0.158 - 0.104  & 0.150 - 0.097  \\
$a_2^q=$ (--0.062) - 0.046&
0.144 - 0.152  & 0.136 - 0.143  & 0.121 - 0.126  & 0.113 - 0.119   \\
$B_2=$ 2.1-7.1&
0.148 - 0.148  & 0.139 - 0.139  & 0.122 - 0.125  &  0.115 - 0.117   \\
\hline
\end{tabular}
\caption{%
Theoretical uncertainty of the form factors at maximal recoil from
each of the parameters.}\label{tbl2}
\end{center}
\end{table}

We then investigate the theoretical uncertainty for the
$B\to\eta^{(\prime)}$ form factors from each of the input
parameters, and the results are presented in Table~\ref{tbl2}. The
form factors $F^{B\eta_q}_{q+,q0,qT}$ in Eq.~(\ref{qfpic}) receive
substantial contributions from the twist-3 distribution amplitudes
$\phi_q^P$ and $\phi_q^T$, which appear together with the mass
ratio $r_q = m_0^q/m_B$. The chiral enhancement scale $m_0^q$
changes rapidly with the mixing angle $\phi$ and with the decay
constants $f_{q,s}$ as illustrated in Fig.~\ref{figm0}. This
explains the sensitivity of the form factors to these inputs. We
notice that $m_0^q$ runs into the unrealistic negative region
easily, as increasing $f_s$, leading to the negative form factor
values in Table~\ref{tbl2}. Therefore, we prefer not to vary
$f_s$. The dependence of the form factors on the Gegenbauer
coefficients $a_2^q$ and $B_2$ is weak, because $a_2^q$ is close
to zero and the gluonic contribution is subdominant. The variation
of the ratios of the gluonic contributions in the
$B\to\eta^{(\prime)}$ form factors with each of the parameters is
listed in Table~\ref{tbl3}. Ignoring the effect from changing
$f_s$ for the reason stated above, we conclude that the gluonic
contribution is negligible, always below 1\%, in the $B\to \eta$
transitions, and may reach order of 10\% in $B\to \eta'$.

\begin{table}[ht]
\begin{center}
\begin{tabular}{l|cccc}\hline
\rule{0mm}{4mm} & $F^{B\eta}_{+,0}(0)$ & $F^{B\eta}_T(0)$ &
$F^{B\eta'}_{+,0}(0)$ & $F^{B\eta'}_T(0)$\\\hline $\omega_B=$
(0.36 - 0.44)&
0.29 - 0.34 &0.25 - 0.30 &2.1 - 2.5 & 1.9 - 2.2  \\
$f_q=$ (1.05 - 1.09)$f_\pi$&
0.75 - 0.21 & 0.73 - 0.18 & 5.7 - 1.5 & 5.6 - 1.3\\
$f_s=$ (1.28 - 1.40)$f_\pi$&
0.14 - (--0.64)&0.12 - (--0.49)&0.94 - (--6.0)&0.81 - (--4.5)\\
$\phi=$ (38.3 - 40.3)&
0.55 - 0.20&0.51 - 0.17&3.7 - 1.6&3.5 - 1.4\\
$m_q=$ (4.0 - 7.2)&
0.24 - 0.37&0.21 - 0.33&1.8 - 2.7&1.6 - 2.5\\
$a_2^q=$ (--0.062) - 0.046&
0.32 - 0.30&0.29 - 0.27&2.4 - 2.2&2.1 - 2.0\\
$B_2=$ 2.1 - 7.1&
0.14 - 0.48&0.13 - 0.43&1.1 - 3.5&0.95 - 3.1
\\\hline
\end{tabular}
\caption{%
Theoretical uncertainty of the ratios (in \%) of the gluonic contributions
in the form factors at maximal recoil from each of the
parameters. }\label{tbl3}
\end{center}
\end{table}

At last, we discuss the impact of the gluonic contribution in the
$B\to\eta^{(\prime)}$ form factors on the predictions for the
$B\to\eta^{(\prime)}K$ branching ratios in QCDF and in PQCD. The
current data of the branching ratios are summarized below
\cite{HFAG}:
\begin{eqnarray}
B(B^\pm\to\eta' K^\pm)&=&(69.7^{+2.8}_{-2.7})\times
10^{-6}\;,\nonumber\\
B(B^0\to\eta' K^0)&=&(64.9\pm 3.5)\times
10^{-6}\;,\nonumber\\
B(B^\pm\to\eta K^\pm)&=&(2.2\pm 0.3)\times 10^{-6}\;,\nonumber\\
B(B^0\to\eta K^0)&<&1.9\times 10^{-6}\;.\label{data}
\end{eqnarray}
Table 4 in the QCDF analysis \cite{BN02} shows the dependence of
the $B\to\eta^{(\prime)}K$ branching ratios on the gluonic
contribution $F_2$: $B(B\to\eta' K)$ increases with $F_2$, but
$B(B\to\eta K)$ decreases. The reason is as follows. The gluonic
contribution enhances the $B\to\eta_q K$ amplitude (containing the
$B\to\eta_q$ transition), such that its cancellation with the
$B\to K\eta_s$ amplitude (containing the $B\to K$ transition)
becomes more exact in the $B\to\eta K$ decays. However, the above
two amplitudes are constructive in the $B\to\eta' K$ decays, whose
branching ratios then exhibit an opposite behavior with the
gluonic contribution. Table 4 in \cite{BN02} also shows the
predictions from the default scenario of inputs (with the strange
quark mass $m_s=100$ MeV), $B(B^\pm\to\eta'K^\pm)\approx 42\times
10^{-6}$ and $B(B^\pm\to\eta K^\pm)\approx 1.7\times 10^{-6}$ for
$F_2=0$, both of which fall short compared to the data. Enlarging
$F_2$, the predicted $B(B^\pm\to\eta'K^\pm)$ increases, but
$B(B^\pm\to\eta K^\pm)$ decreases and deviates more from the
measured value. That is, there is no much room for the gluonic
contribution to play in QCDF. Nevertheless, a smaller $m_s=80$ MeV
does help, since it lifts both $B(B^\pm\to\eta'K^\pm)$ and
$B(B^\pm\to\eta K^\pm)$ as demonstrated in Table 4 of \cite{BN02}.

We have stated that the flavor-singlet amplitudes were not taken
into account in the PQCD study of the $B\to\eta^{(\prime)} K$
decays \cite{KS01}. The predictions $B(B^0\to\eta'K^0)\approx
45\times 10^{-6}$ and $B(B^0\to\eta K^0)\approx 4.6\times 10^{-6}$
were obtained for $m_s=100$ MeV and for the chiral enhancement
scale $m_0^q=1.4$ GeV (see Table 2 of \cite{KS01}), which is
within our parameter range for $m_0^q$ as displayed in
Fig.~\ref{figm0}. Because of the dynamical enhancement of penguin
contributions \cite{KLS,CKL}, both the above branching ratios are
larger than those in QCDF from the default scenario \cite{BN02}.
Comparing the PQCD predictions with the data in Eq.~(\ref{data}),
and knowing their dependence on $F_2$, there is more room for the
gluonic contribution to play apparently.
Though the central value of the gluonic contribution is small, we
can not exclude the possibility of accommodating the observed
$B\to\eta^{(\prime)} K$ branching ratios in PQCD under the current
theoretical uncertainty. The complete PQCD analysis of the
$B\to\eta^{(\prime)} K$ decays including all flavor-singlet
amplitudes will be published elsewhere.

Before concluding, we mention that the form factor
$F_T^{B\eta}(0)=0.16\pm 0.03$ has been derived from light-cone QCD
sum rules \cite{AS0302}, which, however, did not include the
two-parton twist-3 and gluonic contributions. Instead, the
three-parton $\eta$ meson distribution amplitudes were taken into
account. Even so, their result is in agreement with ours in
Tables~\ref{tbl1} and \ref{tbl2}. The form factors
$F_+^{B\eta_q}(0)=-0.023\pm 0.048\;(0.045\pm 0.086)$ and
$F_+^{B\eta_s}(0)=-0.099\pm 0.024\;(-0.066\pm 0.043)$
\cite{WZ0610} have been extracted by fitting parametrization in
the soft-collinear effective theory \cite{BPS05} to the data of
two-body nonleptonic $B$ meson decays, where the numbers in the
parentheses represent the second solution of the fitting. These
values, after considering the large uncertainties, still differ
from our observations dramatically, $F_+^{B\eta_q}(0)=0.190$ and
$F_+^{B\eta_s}(0)=0.0005$. Our $F_+^{B\eta_s}(0)$, containing only
the gluonic contribution at leading order of $\alpha_s$, is much
smaller than $F_+^{B\eta_q}(0)$.

\section{CONCLUSION}

In this paper we have calculated the gluonic contribution to the
$B\to\eta^{(\prime)}$ transition form factors in the large-recoil
region using the PQCD approach. The quark-flavor and singlet-octet
schemes for the $\eta$-$\eta'$ mixing were compared, and it was
found that fewer two-parton twist-3 distribution amplitudes could
be introduced in the former. The leading-twist quark and gluonic
distribution amplitudes of the $\eta_q$ and $\eta_s$ mesons in the
quark-flavor basis were constrained experimentally. The parameters
involved in the two-parton twist-3 quark distribution amplitudes
were determined either by equations of motion associated with the
$\eta$-$\eta'$ mixing, or taken the same as in the pion
distribution amplitudes from QCD sum rules. Therefore, we are able
to predict the unknown gluonic contribution with less theoretical
uncertainty. It has been shown that this contribution is
negligible in the $B\to\eta$ form factors, and reaches few
percents in the $B\to\eta'$ ones. These predictions can be
confronted with the future measurement of the
$B\to\eta^{(\prime)}l\nu$ decay spectra. We have elaborated the
impact of the gluonic contribution on the $B\to\eta^{(\prime)}K$
branching ratios obtained in QCDF and in PQCD. The observation is
that the gluonic contribution does not help accommodating the
measured $B\to\eta^{(\prime)}K$ branching ratios in QCDF, but does
in PQCD.

\vskip 1.0cm We thank C.H. Chen and T. Feldmann for useful
discussions. This work was supported by the National Science
Council of R.O.C. under the Grant No. NSC-95-2112-M-050-MY3 and by
the National Center for Theoretical Sciences of R.O.C.. HNL thanks
Yukawa Institute for Theoretical Physics for her hospitality
during his visit, where this work was completed.

\end{document}